\documentclass{ws-ijmpe}

\def\mib#1{\mbox{\boldmath $#1$}}
\newcommand{\wtilde}[1]{\widetilde{#1}} 
\newcommand{\slp}{\raise.1ex\hbox{$/$}\kern-.63em\hbox{$p$}}
\newcommand{\slk}{\raise.15ex\hbox{$/$}\kern-.53em\hbox{$k$}}
\newcommand{\slpartial}{\raise.15ex\hbox{$/$}\kern-.53em\hbox{$\partial$}}
\newcommand{\ov}{\overline}
\newcommand{\bp}{\mib{p}}

\begin{document}

\markboth{Y.Tsue, H. Fujii and Y. Hashimoto}
{Landau Potential Study of the Chiral Phase Transition in a QCD-Like Theory}


\title{LANDAU POTENTIAL STUDY \\
OF THE CHIRAL PHASE TRANSITION IN A QCD-LIKE THEORY
}

\author{\footnotesize Y.~TSUE
}

\address{Physics Division, Faculty of Science, Kochi University\\
Kochi 780-8520, Japan
\\
tsue@cc.kochi-u.ac.jp}

\author{H.~FUJII}

\address{Institute of Physics, The University of Tokyo\\
Tokyo 153-8902, Japan\\
hfujii@phys.c.u-tokyo.ac.jp}

\author{\footnotesize Y.~HASHIMOTO
}

\address{Department of Applied Science, Kochi University\\
Kochi 780-8520, Japan}

\maketitle

\begin{abstract}
We investigate the chiral phase transition of a QCD-like theory, 
based the shape change of the effective potential near the critical point. 
The potential is constructed with the auxiliary field method,
and a source term coupled to the field is
introduced in order to compute the potential shape numerically.
We also generalize the potential so as to have two independent
order parameters, the quark scalar density and the number density.
We find a tri-critical point locating at $(T,\mu)=(97, 203)$ MeV,
and visualize it as the merging point of three potential minima.
\end{abstract}

\section{Introduction and Summary}

The phase structure of quantum chromodynamics (QCD) at finite temperature 
$T$ and 
quark chemical potential $\mu$ is actively investigated in the context 
of the collider experiments using ultra-relativistic heavy ions,
which create the highly-excited QCD matter.\cite{1} 
The chiral symmetry of QCD, which is spontaneously broken in the vacuum, 
will be restored at sufficiently high temperature and/or quark 
chemical potential. 
We investigate here the chiral phase transition of a QCD-like theory, 
focusing on the shape change of the effective potential near the critical point. 
The QCD-like theory is the renormalization-group (RG) 
improved ladder approximation 
for the Schwinger-Dyson (SD) equation of QCD.\cite{4,5,6} 
This theory describes 
the dynamical chiral symmetry breaking, retaining the correct 
high energy behavior of the quark propagator. Using this theory and 
the low-density expansion, the pion-nucleon sigma term and the quark 
condensate at finite density have been calculated successfully.\cite{7} 

As for the effective potential, 
the Cornwall-Jackiw-Tomboulis (CJT) potential functional\cite{8} has been 
used in several works for the study of dynamical chiral symmetry breaking. 
However, the interpretation of the CJT potential away from the extremum 
is not obvious. 
Thus, in order to investigate the nature of the chiral phase 
transition through the shape change of the effective potential, 
we construct the Landau potential\cite{10} of the 
QCD-like theory\cite{9} 
by using the auxiliary field method, in which the bilocal 
external field is introduced. We generalize the potential
so as to have two independent order parameters,
the quark condensate and the quark number density at fixed 
$T$ and $\mu$. By plotting the 2D contour map in the order parameter plane,
the shape change of the potential near the critical point can be elucidated. 
We find a tri-critical point is located at $(T,\mu)=(97, 203)$ MeV, 
and visualize it as the merging point of three potential minima.

\section{Basic Ingredients of a QCD-Like Theory}

\subsection{In the vacuum}

In the QCD-like theory, first, we neglect the self-interaction between the 
gluons 
and integrate out the gluon fields, leaving the non-local interactions 
between the
quark.
The non-Abelian nature is taken into account in the running of the
coupling  constant.
The generating functional $Z=\int{\cal D}\psi{\cal D}{\bar \psi}{\cal D}A
\exp(i\int d^4x {\cal L}_{\rm QCD})$ is approximated by 
$Z \!\!\sim\!\! \int{\cal D}\psi{\cal D}{\bar \psi}
\exp(i\int d^4x {\cal {\wtilde L}})$ where 
\begin{eqnarray}\label{1}
\int d^4x \mathcal{{\wtilde L}} 
&=&
\int_p\overline{\psi}(p)\ \slp\ \psi(p)
\nonumber\\
& &-\frac{i}{2}\int_{pqk}\!
\psi_\alpha\!\!\left(p-\frac{q}{2}\right)
\overline{\psi}_\beta\!\!\left(p+\frac{q}{2}\right)
K^{\alpha\beta,\gamma\delta}(p,k) 
\psi_\gamma\!\!\left(k+\frac{q}{2}\right)
\ov{\psi}_\delta\!\!\left(k-\frac{q}{2}\right). 
\end{eqnarray}
Here, $\psi(p)$ is the quark field in the momentum space 
and $\int_p\!\!\equiv\!\!\int d^4p/(2\pi)^4$.
The indices, $\alpha, \beta, \cdots$ correspond to the Dirac structure. 
We defined the kernel $K$ as 
\begin{equation}\label{2}
K^{\alpha\beta,\gamma\delta}(p,k)=g^2(\gamma_\mu T^a)^{\delta\alpha}
(\gamma_\nu T^a)^{\beta\gamma}
iD^{\mu\nu}(p-k) \ ,
\end{equation}
where $T^a$ is the color $su(N_c)$ generator and 
$iD^{\mu\nu}(p)$ is the gluon propagator: 
\begin{equation}\label{3}
iD^{\mu\nu}(p)=\frac{g^{\mu\nu}-(1-\alpha)p^\mu p^\nu / p^2}{p^2}
\ .
\end{equation}
Hereafter, we employ the Landau gauge, $\alpha=0$.
The non-Abelian nature of the gluon interaction is treated in this model
as the one-loop running of $g$:
$\ov{g}(\textrm{max}(p_E^2,k_E^2))$ 
with $p_E$ the momentum in Euclidian space.
The divergence of $\ov{g}(p_E^2)$ appearing 
at $p_E=\Lambda_{\rm QCD}$ 
is removed by introducing an infrared 
cutoff parameter $p_{IF}$ as \cite{11} 
\begin{equation}\label{4}
\overline{g}^2(p_E^2)
=\frac{2}{a}\frac{1}{\textrm{ln}((p_E^2+p^2_{IF})/\Lambda_{\rm QCD}^2)} 
\end{equation}
with
$a=1/(8\pi^2)\cdot (11N_c-2N_f)/3$.
Here, 
$N_c(=3)$ and $N_f(=2)$ 
are the numbers of colors and flavors, respectively.

We next introduce the following bilocal auxiliary field 
for non-perturbative analysis:
\begin{equation}\label{5}
\chi_{\alpha\beta}(p,q)=\int_k K^{\alpha\beta,\gamma\delta}(p,k)
\psi_\gamma\left(k+\frac{q}{2}\right)
\ov{\psi}_\delta\left(k-\frac{q}{2}\right).
\end{equation}
This makes the action bilinear in the quark fields $\psi$
and $\ov{\psi}$.
Then, integrating out $\psi$ and
$\ov{\psi}$, we end up with 
$Z=\int {\cal D}\chi \exp i\,\Gamma[\chi]$ where the classical action reads
\begin{eqnarray}\label{6}
\Gamma[\chi]&=&
\frac{1}{2}\int_{pkq}\textrm{tr}[\chi(p,-q)K^{-1}(p,k)\chi(k,p)] 
-i\textrm{Tr Ln}(\ \slp\delta^4(q)(2\pi)^4-\chi(p,-q)) .\ 
\end{eqnarray}
In the mean-field approximation the SD equation is obtained
as the extremum condition for this classical action 
for $\chi$.
The existence of a non-trivial solution for $\chi$ indicates 
the dynamical breaking of the chiral symmetry. 
In the vacuum this solution is expressed as
$\langle \chi_{\alpha\beta} \rangle 
=\Sigma_{\alpha\beta}(p)\delta^4(q)(2\pi)^4$ 
with the mass function $\Sigma$, and 
the effective potential $V[\Sigma]=-\Gamma[\chi]/\int d^4x$ 
becomes
\begin{eqnarray}\label{7}
V[\Sigma]&=&
-\frac{1}{2}\int_{pk}\textrm{tr}[\Sigma(p)K^{-1}(p,k)\Sigma(k)] 
+i\int_p\textrm{tr ln}(\ \slp-\Sigma(p)) .
\end{eqnarray}
Then the SD equation, $\delta V/\delta \Sigma=0$, 
in the improved ladder approximation is written 
\begin{equation}\label{8}
\Sigma_{\alpha\beta}(p)
=\frac{1}{i} \int_k K^{\alpha\beta,\gamma\delta}(p,k) 
\biggl(\frac{1}{\slk -\Sigma(k)} \biggr)_{\gamma\delta} .
\end{equation}
Although the $\Sigma_{\alpha\beta}(p)$ has the general form 
$\Sigma(p^2)\delta_{\alpha\beta}+\Sigma_v(p^2) \slp_{\alpha\beta}$
in the vacuum,
we can set $\Sigma_v(p^2)=0$ in the Landau gauge ($\alpha=0$). 
After carrying out the Wick rotation and the angle integration, 
the SD equation becomes
\begin{eqnarray}\label{9}
\Sigma(p^2_E)&=&
\frac{3C_2(N_c)}{16\pi^2}\int_0^{\infty} k_E^2 
dk_E^2 \overline{g}^2(\textrm{max}(p_E^2,k_E^2)) 
\frac{1}{\textrm{max}(p_E^2,k_E^2)}
\frac{\Sigma(k^2_E)}{k_E^2+\Sigma(k^2_E)^2} \ , \qquad
\end{eqnarray}
where $C_2(N_c)=T^aT^a=(N_c^2-1)/2N_c$.

\subsection{Order parameters}

The quark condensate with the four momentum cutoff $\Lambda$ is defined as  
\begin{eqnarray}\label{10}
\langle \overline{\psi}\psi \rangle_\Lambda&=&
-\frac{1}{i}\int_p\textrm{tr}\left(\frac{1}{\slp-\Sigma(k^2)}\right) 
=-\frac{N_c}{4\pi^2}\int^{\Lambda^2}_0 \!\! dp_E^2 \ 
\frac{p_E^2\Sigma(p_E^2)}{p_E^2+\Sigma(p_E^2)^2}\ . 
\end{eqnarray}
This bare value at the scale $\Lambda$  is converted into 
the value at the lower energy scale $\mu$ (e.g., 1 GeV) 
via the renormalization group equation 
\begin{equation}\label{11}
\langle \overline{\psi}\psi \rangle_{\mu}
= \langle \overline{\psi}\psi \rangle_\Lambda
\left(\frac{\ov{g}^2(\Lambda)}{\ov{g}^2(\mu)}\right)^{1/
[4((12C_2(N_c))^{-1}(11N_c-2N_f)/3)]} \ . 
\end{equation}

Next, the pion decay constant $f_\pi$
is estimated in terms of the mass function 
$\Sigma(p^2)$ by utilizing the Pagels-Stokar formula: 
\begin{equation}\label{12}
f^2_\pi=\frac{N_c}{4\pi^2}\!
\int_0^{\infty}\!\!\!
 dp_E^2 \frac{p_E^2\Sigma(p_E^2)}{(p_E^2+\Sigma^2(p_E^2))^2}\!
\left(\!\!\Sigma(p_E^2)\!-\!
\frac{p_E^2}{2}\frac{d\Sigma(p_E^2)}{dp_E^2}\!\!\right) \ .
\end{equation}
We fix the value of $\Lambda_{\rm QCD}$ 
here 
so as to reproduce the empirical value of $f_\pi$.

\subsection{At finite temperature}

We use the imaginary time formalism to extend the QCD-like
theory to the case with finite temperature and chemical potential, 
making the following replacement:
\begin{equation}\label{13}
\int_p f(p_0,\bp) \longrightarrow 
T\sum_{n=-\infty}^\infty \int\frac{d^3\bp}{(2\pi)^3}f(i\omega_n+\mu,\bp) \ , 
\end{equation}
where $\omega_n=(2n+1)\pi T \; (n \in \mathbb{Z} )$ is the Matsubara 
frequency for the fermion. 
The mass function $\Sigma_{\alpha\beta}(\omega_n,\mib{p})$ 
at finite $T$ and $\mu$ 
decomposes into 
$\Sigma(\omega_n,|\mib{p}|)\delta_{\alpha\beta}
+\Sigma_s(\omega_n,|\mib{p}|)\omega_n(\gamma_0)_{\alpha\beta}
+\Sigma_v(\omega_n,|\mib{p}|)p^i(\gamma_i)_{\alpha\beta}$. 
We assume here $\Sigma_s=\Sigma_v=0$ for simplicity for our 
purpose of demonstrating the usefulness of the effective
potential in the QCD-like theory. 
Further, we use a covariant-like ansatz\cite{11} for the mass function:
\begin{eqnarray}\label{14}
& &\Sigma(\omega_n,\bp) \longrightarrow \Sigma(\hat{p}^2) , 
\end{eqnarray}
where the frequency and the momentum 
appear in the combination $\hat{p}^2=\omega_n^2+|\bp|^2 $
in $\Sigma$.

\section{Landau Potential}

Here we explain a method that we use
to construct the functional form of the effective potential away from the
extremum.
A standard way to assess the potential form is to apply an external
source which is coupled to the field linearly.
It is recognized, however, that using this approach 
we cannot study non-convex potentials,
which is an important feature near the phase transition point.
We thus apply an external source field $J(p,k)$ which is
coupled to the square of the self-energy as 
\begin{equation}\label{18}
\widetilde{V}[\Sigma,J]
=V[\Sigma]+\frac{1}{2}\int_{pk}\textrm{tr}[\Sigma(p)J(p,k)\Sigma(k)] ,
\end{equation}
where $V[\Sigma]$ has been defined in (\ref{7}). 
By imposing the extremum condition for ${\wtilde V}[\Sigma, J]$ 
with respect to 
$\Sigma(p)$, 
we derive the SD equation with the source field $J$ and have a solution 
$\Sigma_J(p)$. 
The effective potential $V$ for the configuration
$\Sigma_J(p)$ is written as 
\begin{eqnarray}\label{20}
{V}[\Sigma_J]
&=&{\widetilde V}[\Sigma_J,J]
-\frac{1}{2}\int_{pk}\textrm{tr}[\Sigma_J(p)J(p,k)\Sigma_J(k)] \ .
\end{eqnarray}
Near the critical points we compute  $V[\Sigma_J]$ 
for a family of $\Sigma_J$ obtained through a particular set of the
function $J(p,k)$. 
Here, among infinite possibilities, we choose one natural choice, 
$J(p,k)= -c K^{-1}(p,k)$, with a parameter $c$,
and study the shape of the effective potential along this variation.

In the numerical evaluation of the
effective potential,
we split $V[\Sigma]$ as 
$V[\Sigma]= (V[\Sigma]-V[0]) + V[0]$.
Here, 
the difference between the free energies $V[\Sigma_J]-V[0]$ is
finite and can be evaluated numerically. 
The last term 
$V[0]\!\!=\!\! i\int_p\textrm{tr ln}\ \slp $ is 
the potential of a free massless quark gas 
and is 
divergent due to the vacuum fluctuations.
In our regularization to remove this divergence, we replace 
the last term $V[0]$ 
with the pressure $P_{\rm{free}}$ of a free massless
quark gas:
$-V[0]\to  P_{\rm{free}} = 
N_cN_fT^4\left[{7\pi^2}/{180}+({1}/{6})\cdot\left({\mu}/{T}\right)^2
+({1}/{12\pi^2})\cdot\left({\mu}/{T}\right)^4 \right]  .
$

In order to introduce another order parameter,
the quark number density $\rho$, 
we first replace the variable $\mu$ with 
the quark number density $\rho$
through the usual Legendre transformation, 
\begin{equation}\label{31}
F(\langle \bar \psi \psi \rangle,\rho, T)
=V(\langle \bar \psi \psi\rangle,\mu,T)+\rho\mu 
\end{equation}
with
$\rho\equiv
-{\partial \, V}(\langle \bar \psi \psi \rangle,\mu,T) /{\partial \, \mu}$.  
Then, by adding a coupling energy with an external potential $\nu$ 
to Eq.~(\ref{31}), 
we define a new effective potential as
\begin{equation}\label{32}
\ov{V}(\langle \bar \psi \psi\rangle, \rho ;\nu,T)
=F(\langle \bar \psi \psi \rangle,\rho, T)-\rho \nu ,
\end{equation}
which may be interpreted as a Landau potential.\cite{NG92}
Note that $\rho$ and $\nu$ are independent variables here.
When
the condition
$\partial \ov{V}/\partial \rho = \mu-\nu = 0$ is imposed,
this new potential  $\ov V$
coincides with the original potential:
$\ov{V}(\langle \bar \psi \psi\rangle, \rho(\mu) ;\mu,T)
=V(\langle \bar \psi \psi\rangle ;\mu,T)$.

\section{Numerical results}

\begin{figure}[b]
\begin{tabular}{cccc}
\resizebox{26mm}{!}
{\includegraphics[height=1cm]{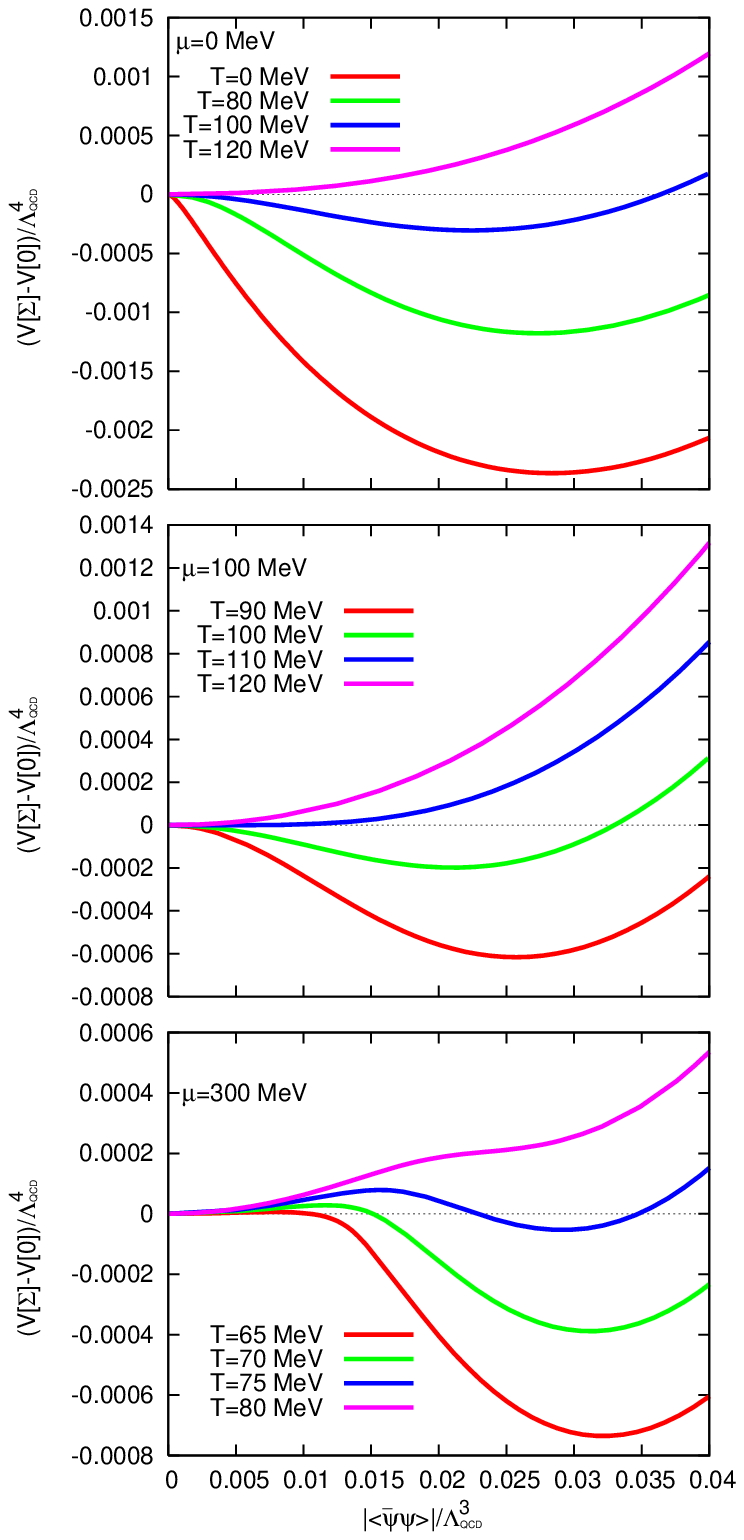}} &
\hspace{-0.8cm}\resizebox{42mm}{!}
{\includegraphics[height=3cm]{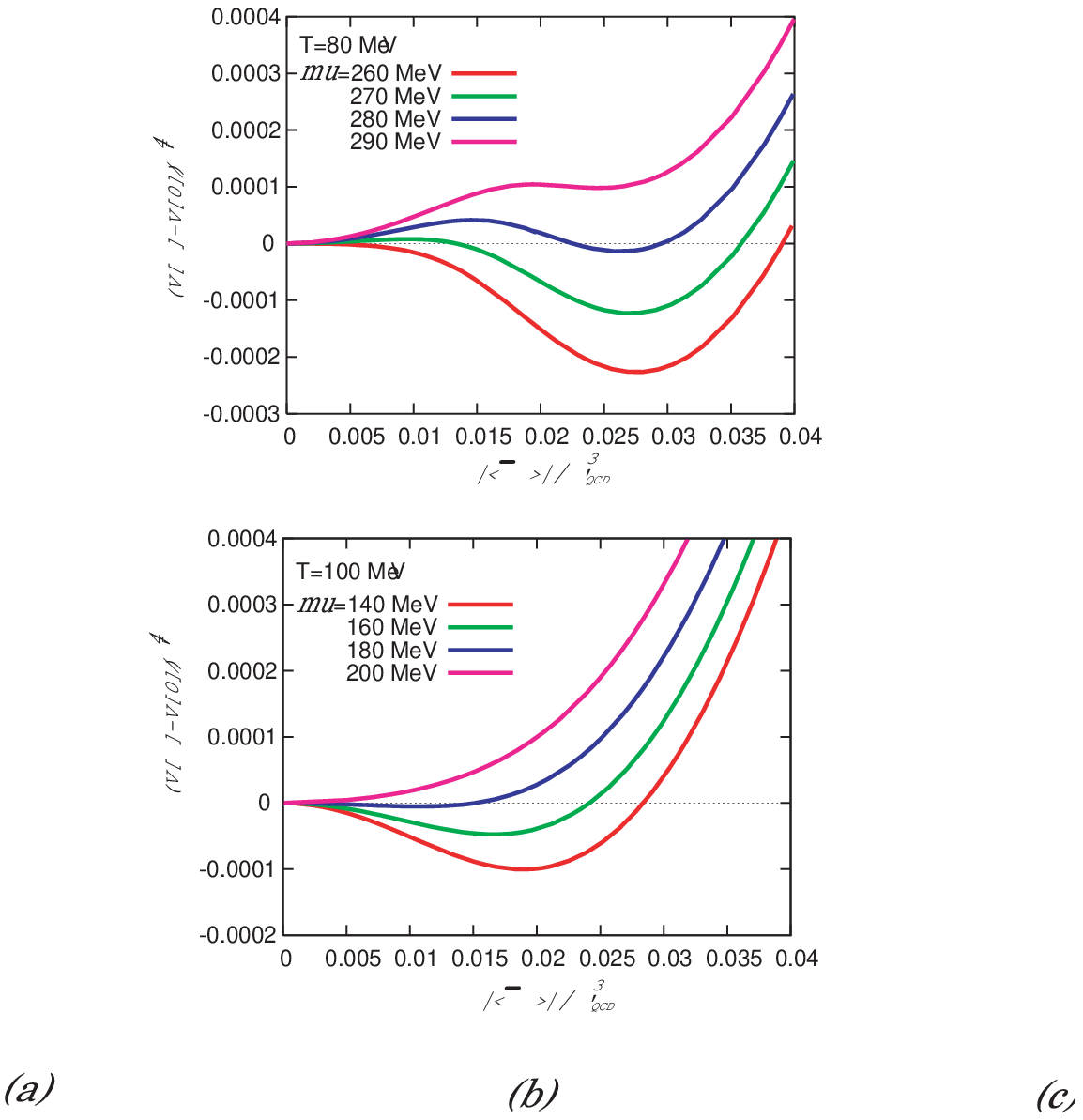}} &
\hspace{-1.2cm}\resizebox{32mm}{!}
{\includegraphics[height=1.2cm]{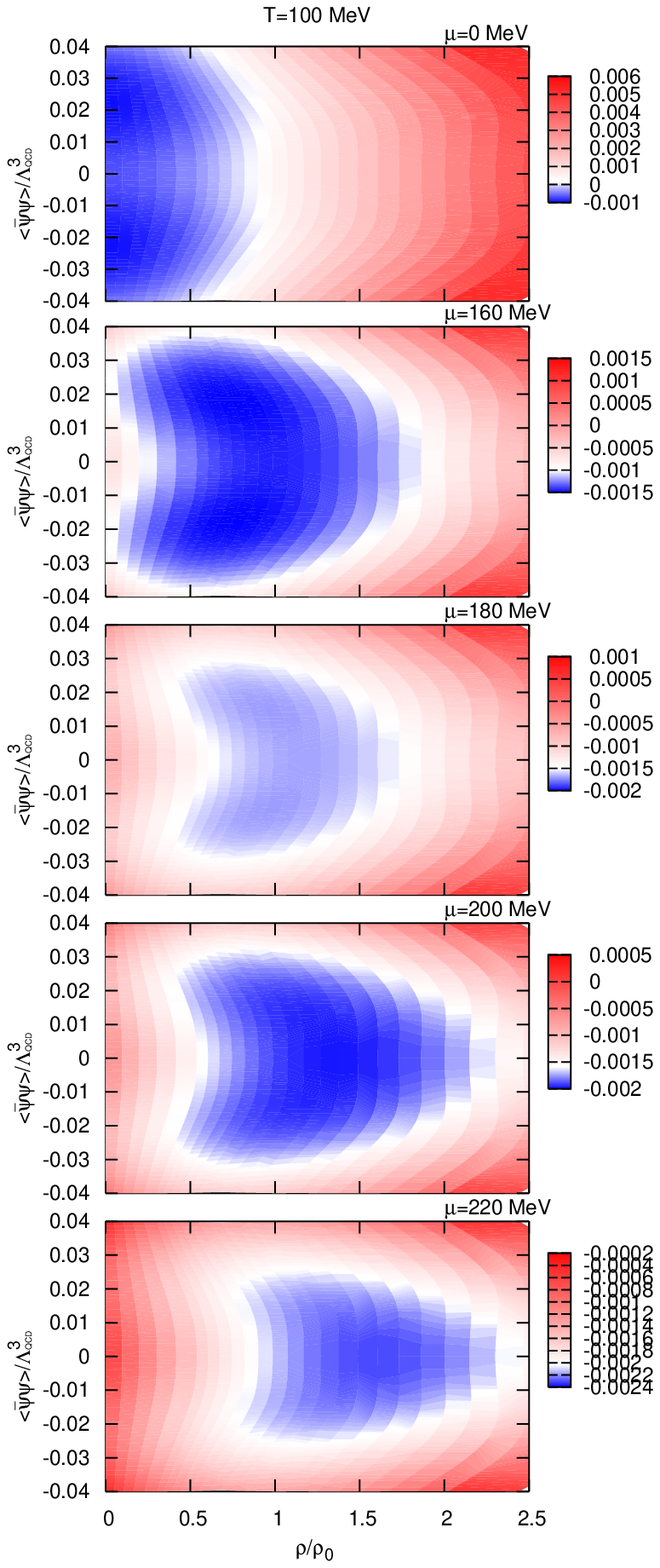}} &
\resizebox{32mm}{!}
{\includegraphics[height=1.1cm]{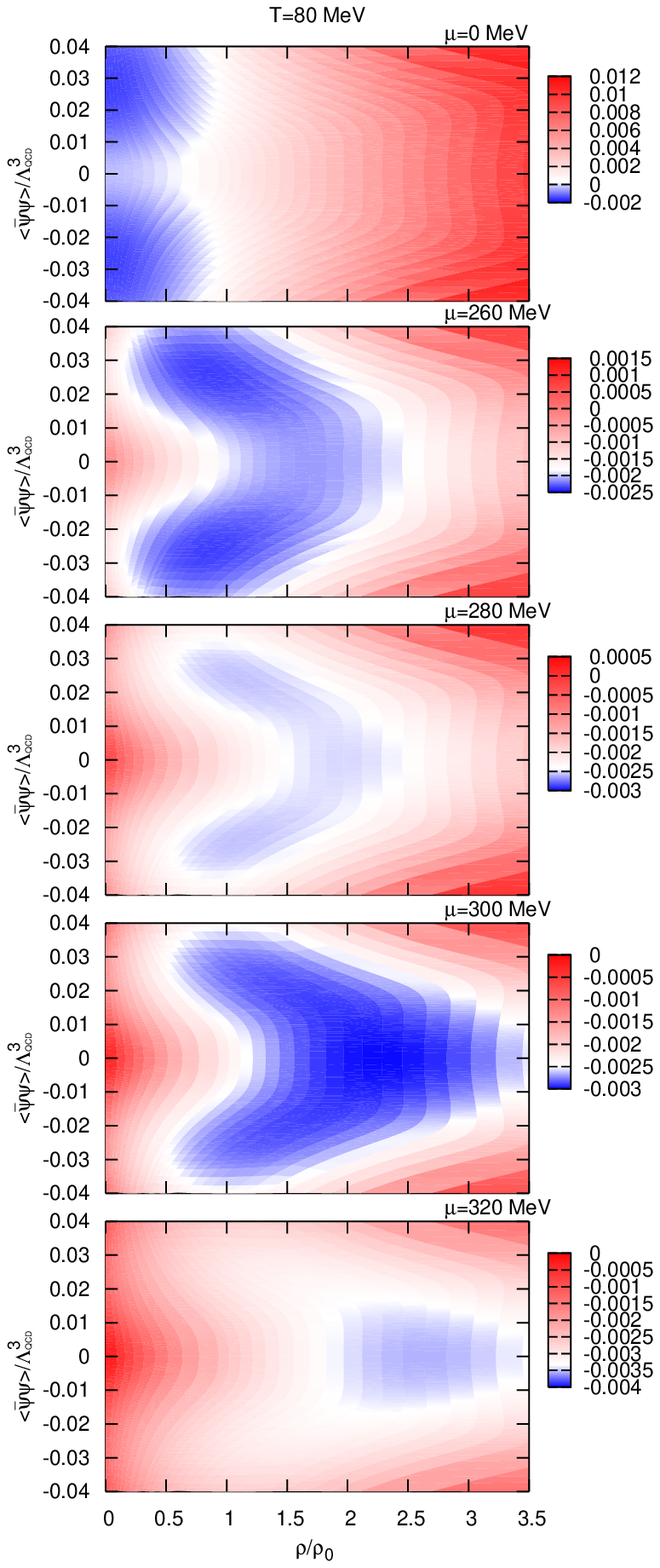}} \\
\end{tabular}
\vspace*{8pt}
\caption{The effective potential, 
$V[\Sigma_c]-V[0]$, with (a) fixed $\mu$ and (b) fixed $T$ 
as a function of 
$|\langle{\bar \psi}\psi\rangle|$. 
(c) The contour map of the Landau potential in the 
$\rho$-$\langle{\bar \psi}\psi\rangle$ plane with various chemical potentials 
for $T=100$ MeV (left) and $T=80$ MeV (right).}
\end{figure}

We plot $V[\Sigma_c]-V[0]$ as a function of 
$|\langle {\ov \psi}\psi\rangle|/\Lambda_{\rm QCD}^3$ 
in Fig. 1(a) at $\mu=0$, 100 and 300 MeV for several values of $T$ 
and in Fig. 1(b) at $T=80$ and 100 MeV with several
values of $\mu$. 
As seen in Fig. 1(a), 
a second-order transition occurs in the range $T=$ 100 -- 120 MeV
in the cases with $\mu=0$ and $\mu=100$ MeV.
At $\mu=300$ MeV, however, a first-order transition occurs 
between $T=75$ MeV and 80 MeV. 
Similarly, in Fig. 1(b), a first-order transition is seen to occur 
when the chemical potential is changed between 
$\mu=280$ MeV and 290 MeV at low temperature ($T=80$ MeV),
while the transition is second order at fixed $T=100$ MeV.

In Fig. 1(c), we plot the contour map of the Landau potential 
given in (\ref{32}) at several values of $\mu$ with fixed $T$, 
namely $T=100$ MeV (left) and $T=80$ MeV (right). 
The vertical axis represents the quark condensate 
$\langle{\bar \psi}\psi\rangle/\Lambda_{\rm QCD}^3$, and the horizontal 
axis represents the quark number density in units of the 
nuclear saturation density $\rho_0$. 
It is shown that 
a second-order transition occurs in the left panels of Fig. 1(c). 
At $\mu=0$, the two minima are positioned symmetrically 
in the $\rho$-$\langle{\bar \psi}\psi\rangle$ plane at $\rho=0$ with
finite values of the quark condensate, $\langle{\bar \psi}\psi\rangle$. 
As the chemical potential increases, the two minima approach each other 
for finite $\rho$, and then fuse continuously to form a single minimum
at $\langle{\bar \psi}\psi\rangle=0$ and finite $\rho$. 
At lower $T$ (in the right panels of Fig. 1(c)), 
a new local minimum appears at $\langle{\bar \psi}\psi\rangle=0$ as $\mu$ 
increases, 
and three local minima exist in a certain range of $\mu$. 
These three minima become energetically degenerated at the critical point,
and we observe a first-order chiral phase transition.
Note that a tri-critical point is located at
$(T_t, \mu_t)\simeq (97 , 203)$ MeV, where these
three minima merge.

\section*{Acknowledgements}

This work was partially
supported by Grants-in-Aid from the Japanese Ministry of Education, 
Culture, Sports, Science and Technology [Nos.15740156 and 18540278 (Y.T.) 
and 13440067 and 16740132 (H.F.)].

\end{document}